# CHARACTERIZATION OF FLEXIBLE RF MICROCOIL DEDICATED TO SURFACE MRI

*M. Woytasik[1], J.-C. Ginefri[2], J.-S. Raynaud[3], M.Poirier-Quinot[2], E. Dufour-Gergam[1], J.-P. Grandchamp[1], L. Darrasse[2], P. Robert[3], J.-P. Gilles[1], E. Martincic[1], O. Girard[2]*

[1]Institut d'Electronique Fondamentale, UMR CNRS 8622, Université Paris Sud, Bât.220, F-91 405 ORSAY Cedex, FRANCE
[2]U2R2M, Université Paris Sud, Bât.220, F-91405 Orsay Cedex, FRANCE
[3]Guerbet Research, 16-24 rue Jean Chaptal, 93 601 Aulnay-Sous-Bois, France

e-mail : marion.woytasik@ief.u-psud.fr

**ABSTRACT**

In Magnetic Resonance Imaging (MRI), to achieve sufficient Signal to Noise Ratio (SNR), the electrical performance of the RF coil is critical. We developed a device (microcoil) based on the original concept of monolithic resonator. This paper presents the used fabrication process based on micromoulding. The dielectric substrates are flexible thin films of polymer, which allow the microcoil to be form fitted to none-plane surface. Electrical characterizations of the RF coils are first performed and results are compared to the attempted values. Proton MRI of a saline phantom using a flexible RF coil of 15 mm in diameter is performed. When the coil is conformed to the phantom surface, a SNR gain up to 2 is achieved as compared to identical but planar RF coil. Finally, the flexible coil is used *in vivo* to perform MRI with high spatial resolution on a mouse using a small animal dedicated scanner operating at in a 2.35 T.

## 1. INTRODUCTION

Currently, the images used to carry out a clinical diagnosis by Magnetic Resonance Imaging (MRI) have a resolution typically limited to $1mm^3$. While this spatial resolution is sufficient for most biomedical applications, an improvement is required when microscopic structures have to be observed (about of $(100\mu m)^3$). This is, for example, the case in NMR imaging of superficial regions such as the skin or the articulations of human extremities. This is also of concern in small animal imaging, which is today in rapid expansion as it allows to develop new diagnosis and therapeutic approaches on models of human pathologies.

The major difficulty to improve the MRI resolution is related to the intrinsic Signal to Noise Ratio (SNR) limitation. This limitation is due to the extreme weakness of the NMR signal and to the thermal noise associated to the radio frequency detection. The usual approach to improve the SNR consists in increasing the intensity of the static magnetic field in order to increase the nuclear magnetization. Unfortunately, this approach leads to the fabrication of complex equipments, based on high-cost magnets and on magnet with reduced bore access.

An alternative approach to improve the SNR consists in decreasing the diameter (d) of the RF coil used to detect the nuclear magnetization. For most biomedical applications involving conducting samples, the noise of the sample is the dominant one as compared to the RF coil noise. In this case, the value of the SNR in the coil proximity depends on its diameter following the scaling law: $SNR \approx d^{-5/2}$ [1]. A diameter reduction of a factor 3 can lead to an increase of SNR of more than one order of magnitude. Small surface RF coils are therefore particularly convenient for local MRI of superficial region or in an intern region when used as implanted RF coil.

However, surface RF coils present a rapid decrease of the sensitivity along the coil axis and in the case of samples with non planar surface, it can provide a significant SNR decrease. The use of flexible coil allowing the matching of the coil to the sample surface can thus improve the magnetic coupling between the sample and the coil and lead to a significant SNR gain.

Flexible RF coils are also interesting in the case of implanted coil, their deformability reducing potential damage against the tissues.

Thus, we developed a fabrication process of RF microcoil compatible with flexible substrates like polymer thin films. To our knowledge, only few teams

*M. Woytasik, J.-C. Ginefri, J.-S. Raynaud, M. Poirier-Quinot, E. Dufour-Gergam,
J.-P. Grandchamp, L. Darrasse, P. Robert, J.-P. Gilles, E. Martincic, O. Girard
Characterization of flexible RF microcoil dedicated to surface IRM*

work on the realization of miniature flexible antenna [2-4] and no imaging using such coil was reported so far.

In this paper, we present the design and the fabrication process of flexible RF microcoils. In order to evaluate the RF sensitivity of such flexible device, we performed both an electrical characterization of the coil performances and a MRI experiment on salin phantom using a microcoil of 15 mm diameter dedicated to 1H imaging at 2.35T (Larmor frequency: 100.1 MHz). Salin phantom is used as a standard solution to simulate the electrical conductivity of biological samples (finger, skin, mouse...). This flexible microcoil was finally used to perform in vivo imaging of a small human mammary tumor induced on a mouse

## 2. RF COIL DESCRIPTION

Conventional surface coils are based on a radio frequency winding, generally in copper, tuned to the NMR frequency of interest by one or more series capacitances. This design becomes problematic to produce RF coils of millimeter size while preserving good electrical performances: the use of a high value capacitor generates a lack of room, setting a size reduction limit and generates magnetic field distortion due to electrical field lines close to the capacitor.

Another approach is to develop a RF microcoil based on the original concept of monolithic resonator.

Such developed resonator consists in two conducting multi-turns circular tracks deposited on both sides of a dielectric substrate (polymer film). Each conducting track is electrically connected to the adjacent one to form a continuous clockwise spiral (Fig.1). This design is an extension of the parallel-plate split conductor principle [5]. The capacitive effect is integrated within the dielectric substrate and the circular tracks set an equivalent inductance ($L_{tot}$).

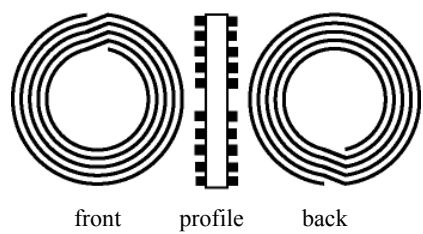

front    profile    back

Figure 1: monolithic resonator configuration

The angular resonance frequency $\omega_0$ is expressed as:

$$\frac{L_{tot}\omega_0}{4Z_0}\tan\left(\frac{\omega_0\sqrt{\varepsilon}\,l_f}{4c}\right)=1 \quad \text{eq(1)}$$

where $l_f$ is the conductor length, $\varepsilon$ the dielectric permittivity and $Z_0$ the characteristic impedance of the conducting line which depends on the dielectric thickness t and the track width w. $L_{tot}$ and $Z_0$ are estimated with a semi-empiric model [5].

Realization using microtechnologies processes enables to manufacture miniature coils. Different RF coils with external diameter in the range of 3 – 15 mm have been designed and realized. Their geometric parameters are presented in Table 1. The 15 and 5 mm diameter RF coils are designed to be used in standard clinical MRI system (static magnetic field ≈ 1 – 2 T) and the 3 mm diameter RF coils are designed to be used at higher static magnetic field (≈ 8 T).

Table 1: RF coil geometric parameters

| External diameter d (mm) | Track width w (μm) | Track spacing s (μm) | Number of turns n |
|---|---|---|---|
| 15 | 400 | 300 | 5 |
|    | 400 | 300 | 4 |
| 5  | 50  | 40  | 20 |
|    | 50  | 20  | 16 |
|    | 60  | 35  | 21 |
|    | 30  | 20  | 14 |
| 3  | 70  | 40  | 7  |

## 3. RF COIL FABRICATION

### 3.1. Substrates

We developed a batch fabrication process of miniature RF coils on polymer thin substrates: polyimide (Kapton® HN, Du Pont De Nemours) and polyether-ether-ketone (PEEK[TM], Victrex). The thicknesses of polymer films were 25, 50 and 125μm.

These materials are thermal and electrical insulators and present good mechanical properties, high chemical resistance, and attractive characteristics of flexibility. Kapton® films have a higher continuous use temperature (320°C) than PEEK[TM] film (260°C). This latter one is a biocompatible polymer and this polymer can withstand many sterilization cycles at high temperature for the medical applications,.

### 3.2. Fabrication process

The fabrication process is based on UV copper micromolding: fabrication of photoresist moulds using UV photolithography and metal electrodeposition through

*M. Woytasik, J.-C. Ginefri, J.-S. Raynaud, M. Poirier-Quinot, E. Dufour-Gergam,*
*J.-P. Grandchamp, L. Darrasse, P. Robert, J.-P. Gilles, E. Martincic, O. Girard*
*Characterization of flexible RF microcoil dedicated to surface IRM*

these moulds thanks to a metallic seed layer previously coated on the substrate.

After the realization of the conducting tracks on the front side of the dielectric substrate, the same process is used for the realization of the conducting tracks on the back face (Fig.2). To ensure the good positioning of the tracks with regard to the already patterned one, an alignment procedure must be performed. Moreover the first side has to be protected during this second step.

The electrodeposited copper presents an electrical resistivity close to that of the bulk value (i.e. 1.7 µΩ.cm). The films are bright and smooth (rms roughness around 10 nm).

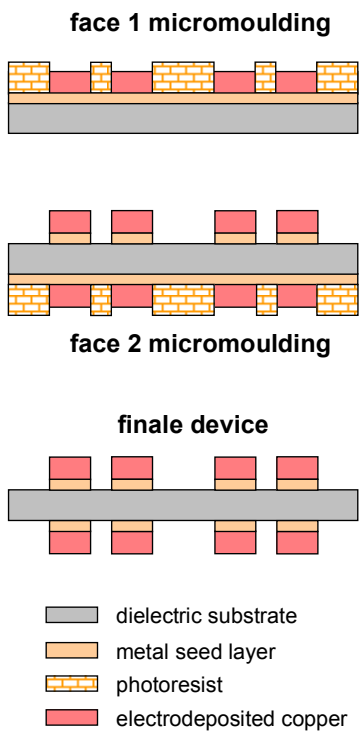

Figure 2: main steps of the RF coil fabrication process

The micromolding process was earlier developed for the fabrication of microcoils on rigid substrates like silicon, glass, LaAlO$_3$ and sapphire [6]. It is compatible with standard integrated circuit (IC) technologies and can be performed by a batch-process.

Some variations of the micromoulding process have being carried out for the realization of coils on flexible substrates. First, in order to correctly perform the contact printing photolithography, polymer substrates must be maintained perfectly flat. For that, they are stuck by a resist on rigid substrates. This method allows an easy separation without damage at the end of the process.

Secondly, in the case of Kapton® substrates, it was necessary to improve the adhesion of the metallic seed layer. The Kapton® surface is modified using an O$_2$ plasma. The improvement in adhesion is caused by a change in the surface topography as well as by the creation of polar groups on the surface [7].

Figure 3 presents some examples of realizations: plane and curved 15 mm diameter RF coils realized on Kapton® film (125 µm) (a) and a batch realization on a 50 µm thick PEEK$^{TM}$ film (2 inches x 2 inches) of 3 and 5 mm diameter RF coils.

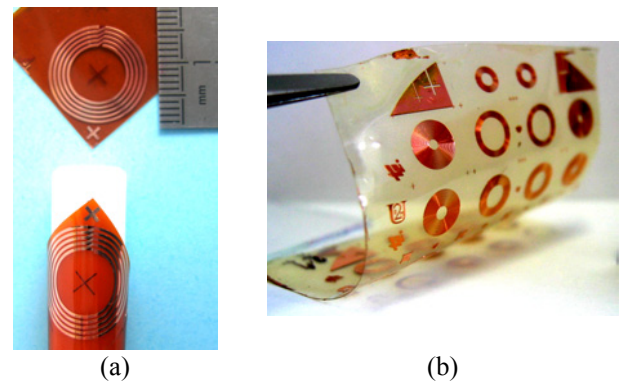

(a) (b)

Figure 3: examples of realization: (a) 15 mm diameter RF coils on a 125 µm thick Kapton® film; (b) 3-5 mm diameter RF coils on a 50 µm thick PEEK$^{TM}$ film

## 4. EXPERIEMENTAL RESULTS

### 4.1 Electrical measurements

The resonance frequency, $f_0$, and the unloaded quality factor, Q, of the RF coils were measured with an inductive method, using a single-loop probe made of a single-turn of copper wire (Ø 0.25 mm)[8]. Its diameter, 3 mm, is smaller than those of the RF coils, allowing an accurate characterization.

Experimental results compared to the theoretical ones are displayed on Fig. 4. The theoretical values are extracted from eq.1

The model used to compute the theoretical resonance frequency, already validated for 15 mm diameter RF coils [6], is also verified in the case of miniaturized coils (diameter less than 5 mm) and for a larger range of frequencies (50 MHz to 400 MHz). For all microcoils, the difference between measured resonance frequencies and theoretical one is less than 10 %.

*M. Woytasik, J.-C. Ginefri, J.-S. Raynaud, M. Poirier-Quinot, E. Dufour-Gergam,*
*J.-P. Grandchamp, L. Darrasse, P. Robert, J.-P. Gilles, E. Martincic, O. Girard*
*Characterization of flexible RF microcoil dedicated to surface IRM*

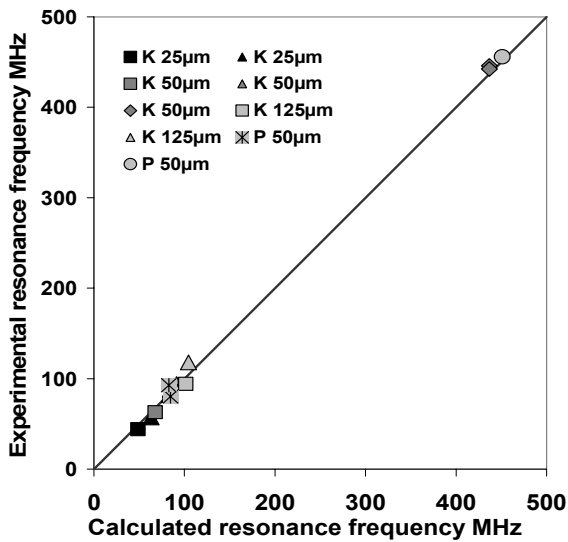

Figure 4: experimental resonance frequency versus theoretical one (eq.1) for RF coil diameters of 15 mm (□), 5 mm (△ and ✷) and 3 mm (◇ and ○) realized on Kapton® (K) and PEEK$^{TM}$ (P)

Q is proportional to the diameter and square root of the resonance frequency of the RF coil. Measured Q values were compared to those predicted by this simple scaling rule applied to a reference coil deposited on Kapton® (50 µm thickness). The reference coil had a Q, diameter, and $f_0$ of 80, 15 mm and 64 MHz of respectively (Fig.5).

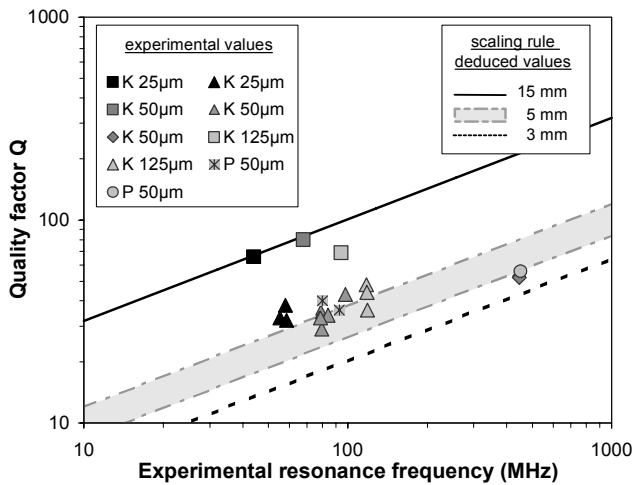

Figure 5: measured and predicted Q respectively represented as a function of experimental $f_0$ for RF coil diameters of 15 mm (□ and solid line), 5 mm (△, ✷ and area) and 3 mm (◇, ○ and dotted line) realized on Kapton® (K) and PEEK$^{TM}$ (P)

The measured Q values are in good agreement with the predicted values (for most of them: difference of less than 10 %), even in the case of smaller diameters, of which design was different from the one of the reference RF coil (n, w, s).

For an equal resonance frequency and for the same RF coil structure (15 mm), the measured Q factor is 25 % lowest compared to that of the rigid RF coil already report [6]. This one was realized on LaAlO$_3$ dielectric (500 µm thick) and presented a Q factor of 109 at 64 MHz. This Q reduction may be likely attributed to the better dielectric performances of LaAlO$_3$ than those of polymer thin films. However, these Q values are sufficient to perform NMR imaging of microscopic structures. And, in the case of samples with none-plane surface, the flexibility of the RF coils may compensate the Q reduction introduced by the use of polymer films improve the spatial resolution. .

### 4.2 MRI characterization

A 4-turn RF coil (15 mm diameter) realized on a Kapton® substrate was used for the proton MRI at room temperature in a static field of 2.35T, which corresponds to a resonance frequency of 100.1 MHz. The RF coil was operated in transmit/receive mode.

*4.2.1. Phantom imaging*

In order to evaluate the detection efficiency of the 15 mm diameter RF coil, preliminary proton NMR imaging experiments have been performed at room temperature on a phantom. This latter one consisted in a 1 cm diameter cylinder filled with a Vistarem® solution (Guerbet Research) at 0.1 mM. Images were obtained using the RF coil in its planar and in its form-fitted configuration. In the second case, the coil is curved to fit the cylindrical shape of the phantom (fig. 6).

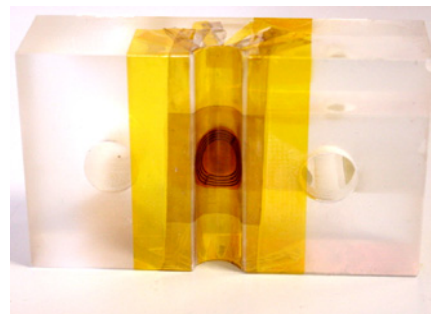

Figure 6: curved RF coil and its support

The obtained phantom images are displayed on figure 7.

*M. Woytasik, J.-C. Ginefri, J.-S. Raynaud, M. Poirier-Quinot, E. Dufour-Gergam,
J.-P. Grandchamp, L. Darrasse, P. Robert, J.-P. Gilles, E. Martincic, O. Girard
Characterization of flexible RF microcoil dedicated to surface IRM*

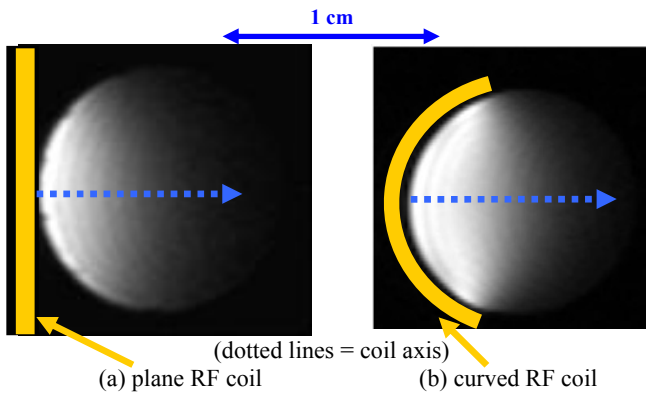

Figure 7: phantom images obtained with a 15 mm RF coil (a) in its plane configuration and (b) and in its deformed one

To evaluate the image quality, the SNR was measured on both images and compared to the one obtained with a standard small animal volume RF coil (10 cm diameter). This latter one achieved a constant SNR over the whole phantom volume, while the SNR obtained with the flexible coil decreased along the coil axis, as expected with a surface RF coil.

As compared to the volume coil, when the flexible RF coil is kept planar, SNR gains of 6 and 2 are achieved at distance inside the phantom of 1 mm and 4 mm respectively. These gains become 8 and 4 when the flexible RF coil is fitted to the cylinder shape. In this case, the SNR decreases more slowly with the distance over the first 5 mm than with the planar RF coil.

*4.2.2. Small animal imaging*

The form-fitted RF coil was used to acquire an in-vivo image of a subcutaneous tumor model induced in a mouse. The image was performed with a spatial resolution of 1000 x 150 x 78 $\mu m^3$ in 17 min (Fig.8).

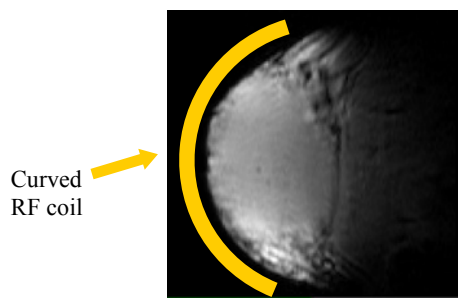

Figure 8: MRI of a subcutaneous tumor model induced in a mouse acquired in vivo with the curved RF coil

The whole tumor is clearly distinguished and many well-defined internal structures are detectable.

To our knowledge, this result forms the first in-vivo NMR image obtained with a form-fitted miniature RF coil realized by a microtechnology process.

## 5. CONCLUSION

In this paper, we proposed a fabrication process of flexible RF microcoil based on micromoulding steps. These coils, 3 to 15 mm in external diameter, are developed for NMR microsopy and are based on the original concept of monolithic resonator. Their resonance frequencies and their unloaded quality factors are in good agreement with predicted ones (less than < 10%) and are sufficient to performed microscopy MRI experiments.

The first MRI experiments demonstrate that, when the coil is form-fitted to the sample surface, a SNR gain is achieved in comparison to a non deformed coil.

However, the use of a miniature RF coil involves a reduction of the field of sight covered by the coil. This drawback could be overcome by associating the signals coming from a network of adjacent micro-antenna. The choice of an optimal configuration depends on the application.